\documentclass[twocolumn,english,aps,prl,10pt,superscriptaddress]{revtex4}
\usepackage[T1]{fontenc}
\usepackage{amsmath,graphicx,amssymb,epsfig,babel,dsfont,color}

\begin{document}

\title{Heat transfer between two metals through subnanometric vacuum gaps}

\author{Riccardo Messina}
\affiliation{Laboratoire Charles Fabry, UMR 8501, Institut d'Optique, CNRS, Universit\'{e} Paris-Saclay,
2 Avenue Augustin Fresnel, 91127 Palaiseau Cedex, France}

\author{Svend-Age Biehs}
\affiliation{Institut f\"{u}r Physik, Carl von Ossietzky Universit\"{a}t, D-26111 Oldenburg, Germany}

\author{Till Ziehm}
\affiliation{Institut f\"{u}r Physik, Carl von Ossietzky Universit\"{a}t, D-26111 Oldenburg, Germany}

\author{Achim Kittel}
\affiliation{Institut f\"{u}r Physik, Carl von Ossietzky Universit\"{a}t, D-26111 Oldenburg, Germany}

\author{Philippe Ben-Abdallah}
\email{pba@institutoptique.fr} 
\affiliation{Laboratoire Charles Fabry, UMR 8501, Institut d'Optique, CNRS, Universit\'{e} Paris-Saclay,
2 Avenue Augustin Fresnel, 91127 Palaiseau Cedex, France}

\date{\today}

\begin{abstract}
We theoretically investigate the heat transfer between two metals across a vacuum gap in extreme near-field regime by quantifying the relative contribution of electrons, phonons and photons. We show that electrons play a dominant role in the heat transfer between two metals at subnanometric distance subject to a temperature gradient. Moreover, we demonstrate that this effect is dramatically amplified in the presence of an applied bias voltage. These results could pave the way to novel strategies for thermal management and energy conversion in extreme near-field regime. 
\end{abstract}

\maketitle

The transition from radiation to conduction regimes between two bodies at different temperatures when their separation distance is reduced to subnanometric gaps is an emerging problem in physics \cite{Volz,Chen}. Recent experimental works carried out by two different teams \cite{Kittel,Reddy} have explored this problem by making a direct measurement of heat power exchanged between a metallic tip and a metallic plate separated by vacuum gaps of angstrom to nanometer width. Nevertheless, these experiments come to radically different conclusions. On one hand, Kittel et al. have reported a strong deviation between their experimental results and the predictions coming from Rytov's fluctuational electrodynamic theory (FED)~\cite{Rytov,Polder}. On the other hand, Reddy et al. found a relatively good agreement with this theory down to few-angstrom separation distances. Unfortunately, due to the lack of a complete theory to describe the multichannel energy exchange at this scale, this problem still remains open. The two experimental setups are similar. They are based on custom-fabricated Scanning Thermal Microscope (SThM) gold probes above a substrate in an ultra-high vacuum environment. In Kittel's experiment the extreme end of the probe can be modeled as a sphere with a radius of curvature of about $30\,$nm, while in Reddy's experiment the tip has a radius of $150\,$nm. Moreover, in the first experimental setup heat transfer takes place at cryogenic temperatures ($T_\text{probe}=280\,$K and $T_\text{sample}=120\,$K), while in the second setup around ambient temperature ($T_\text{probe}=303\,$K and $T_\text{sample}=343\,$K).

In this Letter we explore fundamentally heat transfer in extreme near-field regime by considering the contribution of all possible channels: electrons, phonons and photons. We first describe the transfer between two bulk metallic samples in plane--plane geometry when the separation distance is reduced from nanometer down to angstrom gaps. In particular, we discuss quantitatively and qualitatively the role played by each carrier on the heat transfer in this range of distances, showing that electrons play a major role when approaching the contact. We also show how heat transfer is dramatically affected by the presence of an applied bias voltage. Next, we develop a simple model to evaluate the heat transfer between a SThM tip and a sample and compare our theoretical predictions with recent experimental results.

\begin{figure}[!h]
	\centering
	\includegraphics[scale=0.3,angle=0]{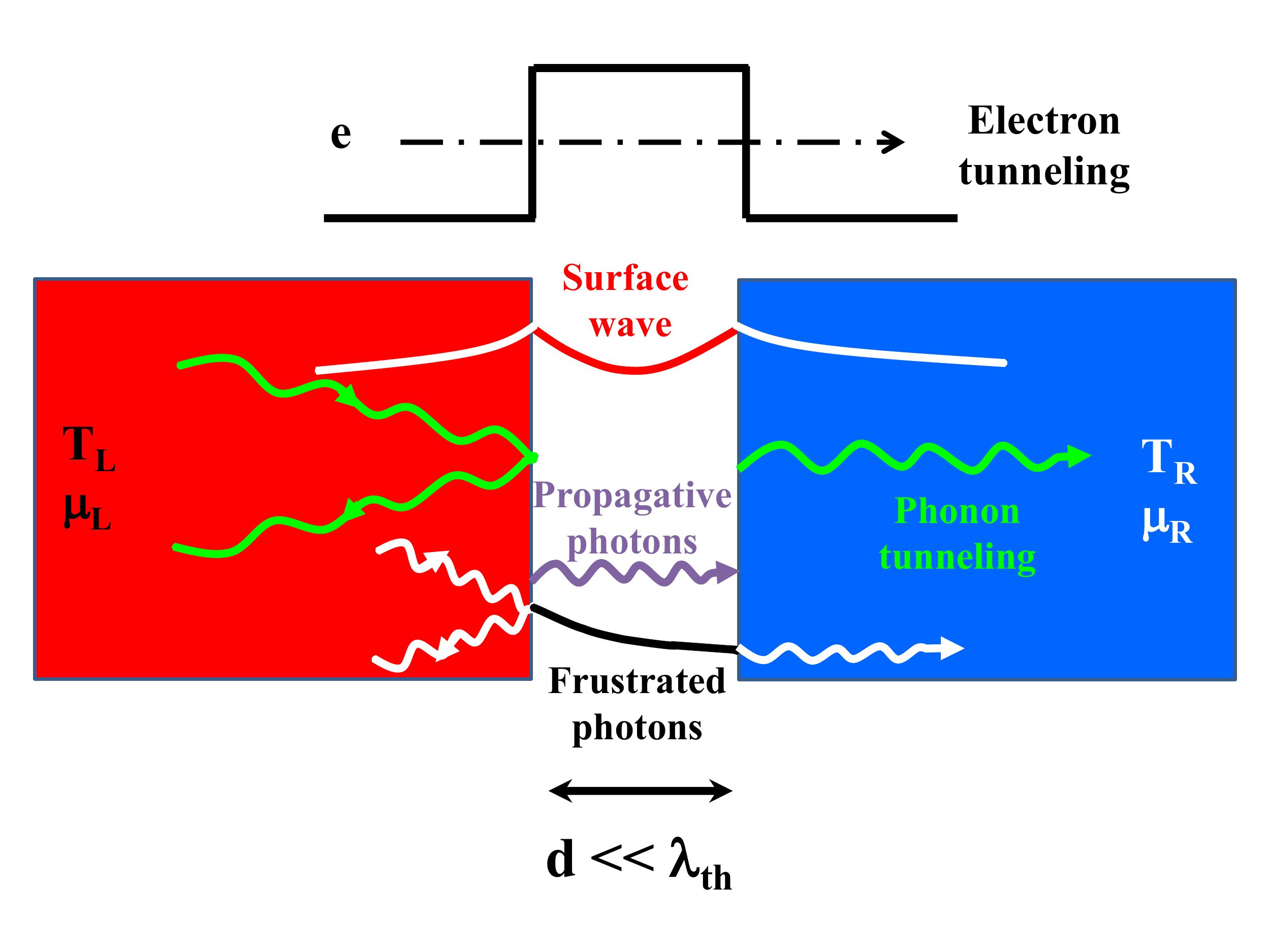}
	\caption{Two solids out of thermal equilibrium separated by a vacuum gap of thickness $d$ exchange heat. When the separation distance $d$ is of the order of the thermal wavelength $\lambda_\text{th}$, heat transfer is mainly due to evanescent photons. When $d$ is much smaller than $\lambda_\text{th}$ (typically at subnanometric distances for reservoir temperatures around the ambient temperature) and becomes even smaller than a nanometer, electrons and acoustic phonons contribute to the heat transfer by tunneling effect through the vacuum gap.} 
	\label{Fig_1}
\end{figure}

We start by considering a system (see Fig.~\ref{Fig_1}) made of two reservoirs at fixed temperature and chemical potential $T_{L,R}$ and $\mu_{L,R}$ separated by a vacuum gap of thickness $d$. Neglecting, in first approximation, the coupling between electrons, phonons and photons, the heat flux transferred from the left ($L$) to the right reservoir ($R$) can be decomposed into three contributions
\begin{equation}
J_h=J_h^\text{(el)}+J_h^\text{(ph)}+J_h^\text{(rad)}
\end{equation}
respectively due to electrons (el), acoustic phonons (ph) and photons (rad). We stress that the contribution of optical phonyons is taken into account in the photonic contribution $J^\text{(rad)}$ through the dielectric permittivity describing the optical response of each material. We detail below the calculation of each of these contributions, starting with the electronic flux by tunneling effect. This flux can be calculated once the effective potential barrier describing the vacuum gap is estimated. In the case of a simple rectangular barrier, the transmission probability $\mathcal{T}(E_x)$ of electrons of normal energy $E_x$ through this barrier reads~\cite{Tannoudji}
\begin{equation}
\mathcal{T}(E_x)= \frac{4E_x(E_x-V)}{4E_x(E_x-V)+V^2\sin^2\bigl(k_{2x}(E_x,V)d\bigr)},
\label{Landauer_elec_coeff}
\end{equation}
where $k_{2x}(E_x,V)=\sqrt{2 m_e(E_x-V)}/\hbar$ denotes the normal components of wavectors inside the gap, $m_e$ being the electron mass and $\hbar$ the reduced Plank constant. We emphasize that the transmission probability $\mathcal{T}(E_x)$ also implicitly depends on the distance $d$ and barrier height $V$. The latter is written here as $V(d)=V_\text{eV}(d)+E_F$, i.e. the sum of the Fermi energy $E_F$ ($E_F=5.53\,$eV for gold) and the distance-dependent potential $V_\text{eV}(d)$, for which the data taken from \cite{Kiejna} have been fitted from DFT calculations with the log-scale law $V_\text{eV}(d)=V_0 \ln(1+d/1\text{\AA})$, where $V_0=1.25\,$eV. This result can be easily generalized to the case of a spatially dependent potential barrier~\cite{Zhang}
\begin{figure}
\centering
\includegraphics[angle=0,scale=0.26]{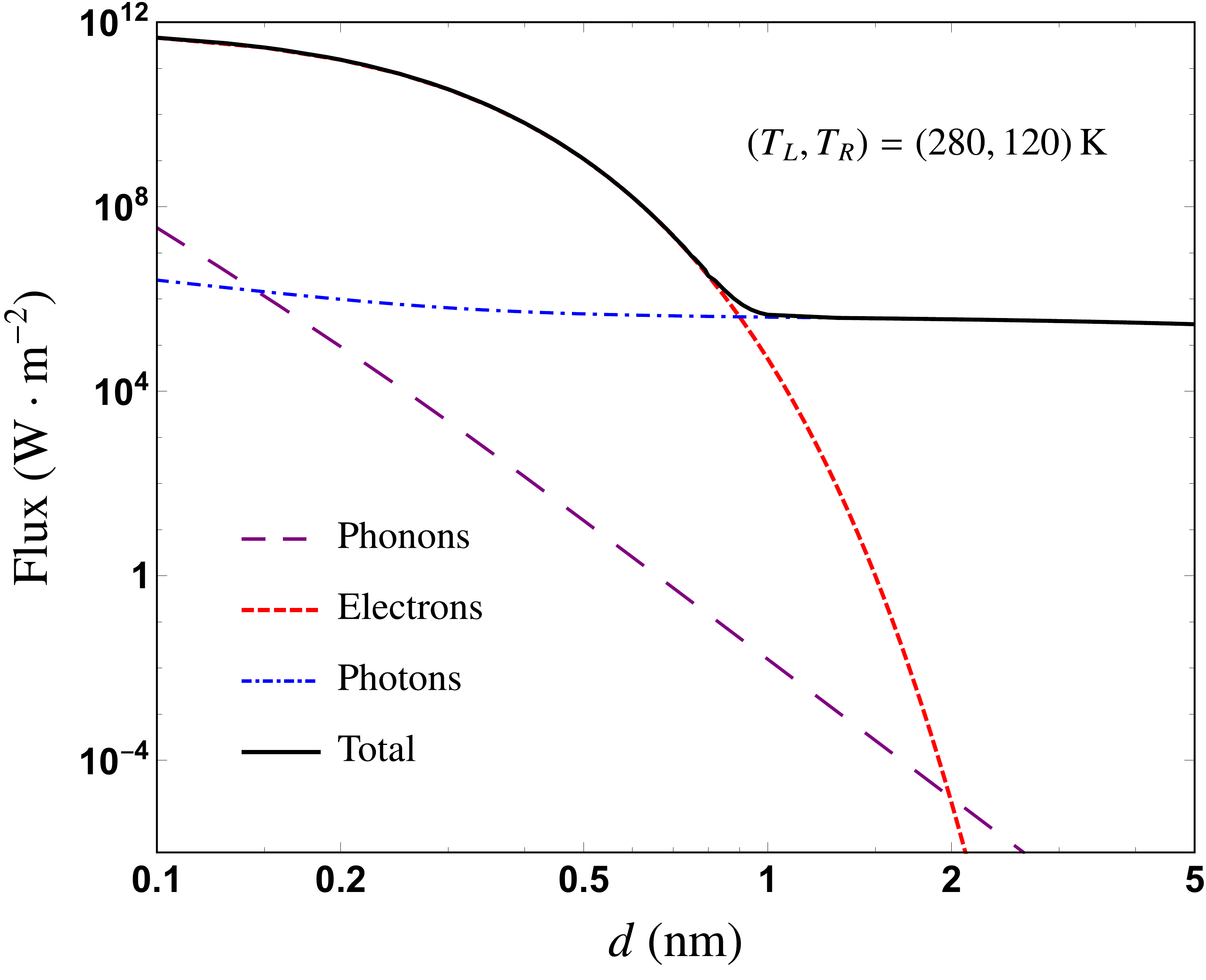}
\caption{Heat transfer at subnanometric scale between two gold samples in plane--plane geometry. The temperatures are $T_L=280\,$K and $T_R=120\,$K and no bias voltage is applied.} 
\label{Fig_2}
\end{figure} 

It follows that the heat flux carried by the electrons through tunnel effect can be calculated by summing over all energies $E_x=\frac{1}{2}m_e v_x^2$ in the $x$-direction normal to the surface. Following Simmons~\cite{Simmons}, the net electronic heat flux reads
\begin{equation}
J_{h}^\text{(el)}=\int_{0}^{\infty}dE_x E_x[N_L(E_x,T_L)-N_R(E_x,T_R)] \mathcal{T}(E_x),
\label{elec_flux}
\end{equation}
where $N_i(E_x,T_i)dE_x$ ($i=L,R$), with $N_i(E_x,T_i)=\frac{m_e k_B T_i}{2\pi^2\hbar^3}\ln[1+\exp(-\frac{E_x-E_F-\mu_i}{k_B T_i})]$, denotes the number of electrons in the reservoir $i$ with a normal energy between $E_x$ and $E_x+dE_x$ across a unit area per unit time. Electrons above (below) the chemical potential of each reservoir contribute positively (negatively) to the flux. Similarly the electric current density reads 
\begin{equation}
J_{e}=e\int_{0}^{\infty}dE_x[N_L(E_x,T_L)-N_R(E_x,T_R)] \mathcal{T}(E_x).
\label{elec_current}
\end{equation}

The heat flux carried by conduction is associated with acoustic-phonon tunneling. Recent theoretical works have quantified the contribution of this channel for exchanges between piezoelectric materials~\cite{Prunnila}, polar materials~\cite{Chen}, semiconductors~\cite{Sellan} and metals~\cite{Pendry} and compared it with near-field heat exchanges. However, to date its relative contribution with respect to the electronic channel has not been addressed. In the long-wavelength approximation this energy transfer can be calculated using the continuous elastic solid theory. In this case, the net heat flux exchanged between two reservoirs reads
\begin{equation}
J_{h}^\text{(ph)}=\underset{l=L,T}{\sum}\int_{0}^{+\infty}\!\frac{d\omega}{2\pi} \hbar\omega \Delta n \int_{0}^{k_c}\frac{d\kappa}{2\pi}\kappa\,\mathcal{T}_l^{(ph)}(\omega,\kappa),
\label{flux_phonon}
\end{equation}
$\kappa$ being the parallel component of phonons wavector, $\kappa_c=\pi/a$ its cutoff ($a$ being the lattice constant) and $\Delta=[n(\omega,T_L)-n(\omega,T_R)]$ the difference of two distribution functions $n(\omega,T_i)=[\exp(\hbar\omega/k_B T_i)-1]^{-1}$ at $T_i$ $(i=L,R)$. The transmission coefficients for the energy of longitudinal ($L$) and transversal ($T$) phonons read~\cite{Pendry}
\begin{equation}
\mathcal{T}_L^\text{(ph)}=\frac{|T_{LL}|^2 c_L^2| k_{Lx}|+| T_{TL}|^2 c_T^2| k_{Tx}|}{c_L^2| k_{Lx}| },
\end{equation}
\begin{equation}
\mathcal{T}_T^\text{(ph)}=\frac{| T_{LT}|^2 c_L^2| k_{Lx}|+| T_{TT}|^2 c_T^2| k_{Tx}|}{c_T^2| k_{Tx}| },
\end{equation}
where $k_{lx}$ ($l=L,T$) denotes the normal component of wave vector (with $k_{lx}+\kappa^2=k_l^2$, $k_l=\omega_l/c_l$ being the modulus of wavector at the frequency $\omega_l$ and $c_l$ the velocity of $l-$phonons), while $T_{lq}$ ($l,q=L,T$) are the transmission coefficients of phonons across the gap (defined in Ref.~\cite{Pendry}) which are directly related to the elastic properties of the solid and to its lattice constants. 

\begin{figure}
	\centering
	\includegraphics[angle=0,scale=0.27]{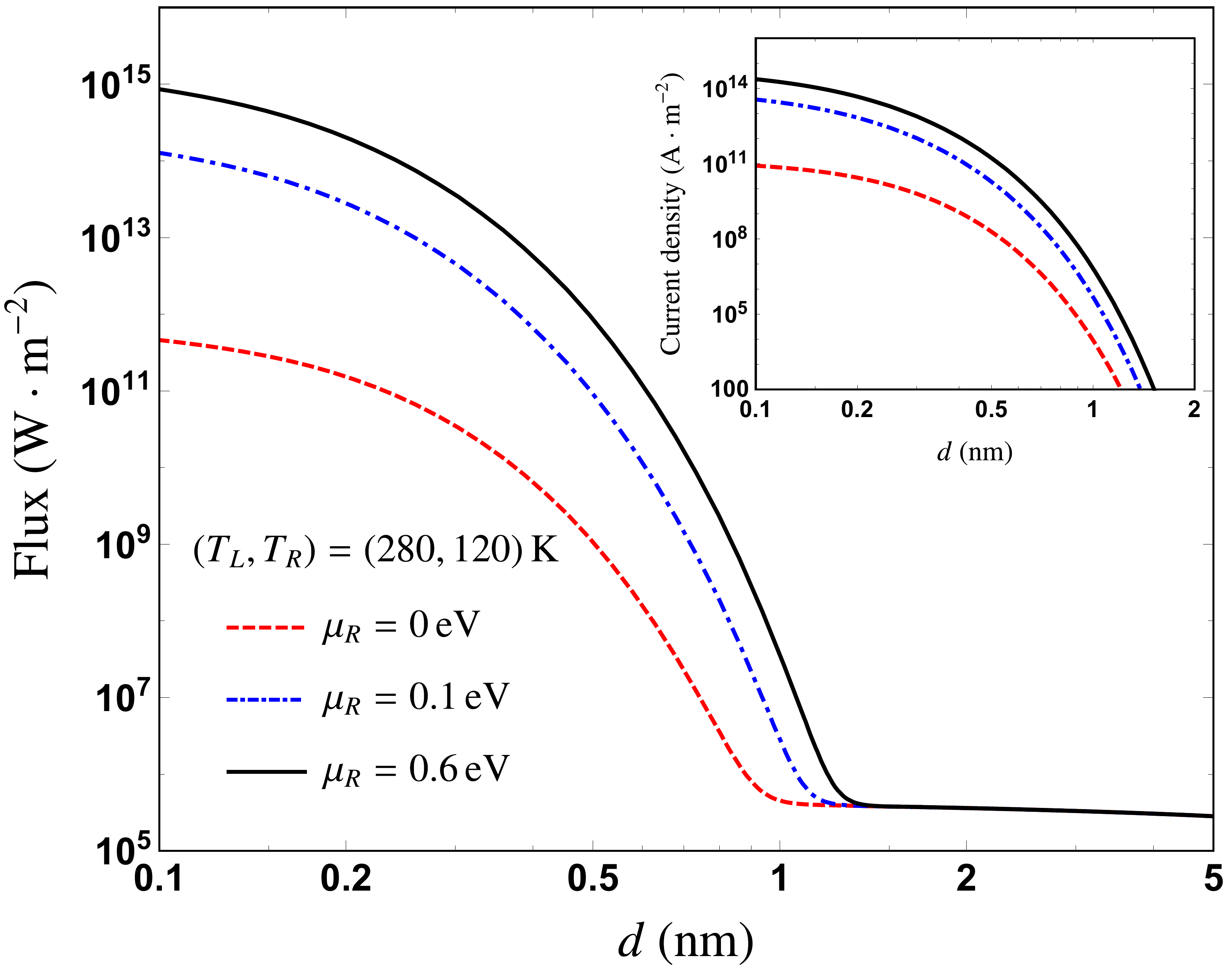}
	\caption{Influence of a bias voltage on the heat flux between two gold samples in plane--plane geometry. The temperatures are $T_L=280\,$K and $T_R=120\,$K. The main part (inset) of the plot shows the heat flux (current density) for three different values of the chemical potential $\mu_R$, assuming $\mu_L=0$.} 
	\label{Fig_bias}
\end{figure} 

As for the flux carried by photons, it can be evaluated from FED using the Polder and van Hove formalism~\cite{Polder} by using a nonlocal effects~\cite{Ford}. The net exchanged heat flux reads
\begin{equation}
J_h^\text{(rad)}= \sum_{p=TE,TM}{\sum}\int_0^\infty\!\frac{d\omega}{2\pi}\ \hbar\omega \Delta n \int_{0}^{\infty} \frac{d\kappa}{2 \pi} \kappa\, \mathcal{T}_p^\text{(rad)}(\omega,\kappa), 
\label{Eq:Flux_D}
\end{equation}
where $\mathcal{T}_p^\text{(rad)}(\omega,\kappa)$ denotes the energy transmission coefficient
\begin{equation}
\begin{split} 
\mathcal{T}^\text{{(rad)}}_p(\omega,\kappa)&= \begin{cases} {\displaystyle \frac{(1-|\rho_{L,p}|^2)(1-|\rho_{R,p}|^2) }{|1- \rho_{L,p}\,\rho_{R,p}\,e^{2i k_x d}|^2}}, & \kappa < \frac{\omega}{c},\\
{\displaystyle \frac{4\, \text{Im}\left(\rho_{L,p}\right)\text{Im}(\rho_{R,p})e^{- 2 \text{Im}(k_x) d}}{|1- \rho_{L,p}\,\rho_{R,p}\,e^{- 2 \text{Im}(k_x) d}|^2}}, & \kappa > \frac{\omega}{c}, \end{cases}\\
\label{transmission_coefficients}
\end{split} 
\end{equation}
for the two polarizations $p = TE,TM$, taking into account the contributions of propagating ($\kappa<\omega/c$, $\kappa$ being the component of the wave vector parallel to the slabs) and evanescent waves ($\kappa>\omega/c$). $\rho_{L,p}$ and $\rho_{R,p}$ are the reflection coefficients of the two reservoirs and $k_{x} = \sqrt{\omega^2/c^2 -\kappa^2}$ is the normal component of the wave vector inside the vacuum gap. When nonlocal effects are neglected and the reservoirs are made of bulk gold, the reflection coefficients reduce to the Fresnel coefficients defined with the dielectric permittivity
$\epsilon(\omega)=1-\omega_p^2/\omega(\omega+i\gamma)$ ($\omega_p=13.71\times 10^{15}\,$rad\,s$^{-1}$ being the plasma frequency and $\gamma=4.05\times 10^{13}\,\text{s}^{-1}$ the damping coefficient). If the nonlocal scenario, the Fresnel coefficients must be replaced by reflections coefficients defined in terms of surface impedances~\cite{Ford}. In the following, our results for the photonic contribution to heat flux always include nonlocal effects.

We emphasize that the expressions given above are valid for any choice of temperatures and voltage bias. In the linear-response approximation (for small $\Delta T=T_L-T_R$ and $\Delta \mu=\mu_R-\mu_L$) the electric and heat currents $J_e$ and $J_h$ are linearly related to the thermodynamic forces $\mathcal{F}_e=\Delta V/T$ (where $\Delta V=\Delta \mu/e$) and $\mathcal{F}_h=\Delta T/T^2$ by the Onsager relations~\cite{de Groot}
\begin{equation}
\left(\begin{array}{c}
J_{e}\\
J_{h}
\end{array}\right)=\left(\begin{array}{cc}
L_{ee} & L_{eh}\\
L_{he} & L_{hh}
\end{array}\right)\left(\begin{array}{c}
\mathcal{F_{\mathrm{e}}}\\
\mathcal{F_{\mathrm{h}}}
\end{array}\right),\label{Eq:Onsager}
\end{equation}
where $L_{a,b}$ $(a,b=e,h)$ are the Onsager coefficients which are related to the familiar transport coefficients. These fluxes can be calculated using the general expressions we have introduced. 

Fig.~\ref{Fig_2} shows the net heat flux exchanged between two gold parallel samples with respect to their separation distance in the range from 1\,\AA\, to 5\,nm in a double logarithmic plot. The temperatures $T_L=280\,$K and $T_L=120\,$K correspond to the parameters of the experiment~\cite{Kittel}. Above the nanometer we observe that the heat exchange is mainly driven by evanescent photons (for the temperatures chosen here the blackbody limit $\phi_{BB}=\sigma(T_1^4-T_2^4)$ is around $300\,$W$\cdot$m$^{-2}$). Below one nanometer the contribution of photons is slightly enhanced by nonlocal effects (Landau damping). A saturation sets in for a distance on the order of the Thomas-Fermi length. Around $d=1 \,$nm the electronic contribution becomes comparable with the photonic part and finally dominates heat exchanges at subnanometric distances. We also notice that the contribution of phonons, which scales as $d^{-10}$, becomes also more important than the one of photons at a few angstrom separation distance. This confirms the conclusions obtained by previous works~\cite{Prunnila, Pendry}. However, the magnitude of the flux carried by elastic waves remains several orders of magnitude smaller than the flux associated to electronic tunneling. We also stress that, while the contribution of the three carriers discussed above quantitatively depend on the choice of the two temperatures $T_L$ and $T_R$, the main features discussed here, namely the leading role of electronic flux at subnanometric distances and the transition around 1\,nm between mainly photonic and mainly electronic flux, are basically unaffected for temperatures close to ambient temperature.

\begin{figure}
	\centering
	\includegraphics[angle=0,scale=0.38]{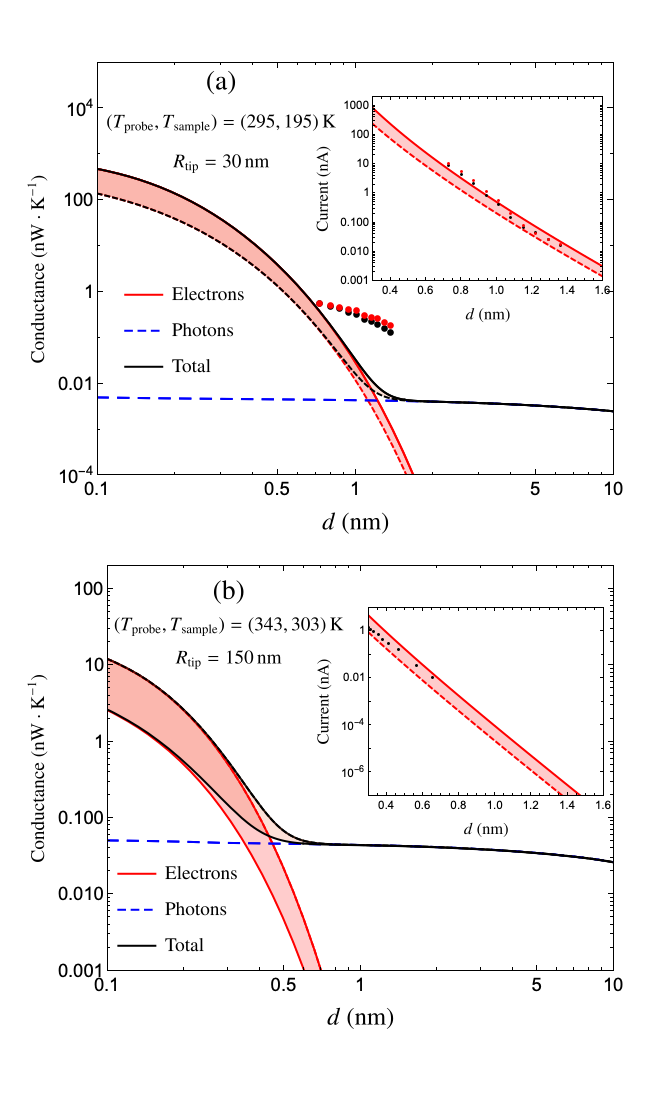}
	\caption{Thermal conductance at subnanometric distance in a SThM geometric configuration made of a Au probe above a Au sample in (a) Kittel's experimental conditions ($T_\text{probe}=295\,$K, $T_\text{sample}=195\,$K and $R=150\,$nm) and (b) Reddy's experimental condition ($T_\text{probe}=343\,$K, $T_\text{sample}=303\,$K and $R=30 \,$nm). Inset: tunneling current as a function of distance $d$. The bias voltage taken in theoretical results is 600\,mV in (a), 8\,mV in (b). The points correspond to experimental data.} 
	\label{Fig_exp}
\end{figure} 

It is now important to address the issue of the impact of the presence of an applied bias voltage between the two bodies. This situation corresponds for example to Kittel's experimental setup where a bias voltage of $600\,$mV is applied between the tip and the sample. The presence of a non-vanishing voltage modifies the electronic chemical potential $\mu$, thus their distribution function. In Fig.~\ref{Fig_bias} we show the value of the thermal power and current exchanged between two gold planar samples in the temperature conditions Kittel's experiment with respect to the separation distance for different values of chemical potential. We observe that below one nanometer both the total power and the current are increased by several orders of magnitude because of this external voltage. This enhancement is of the order of three orders of magnitude at 1\,\AA\ separation distance, with an applied bias voltage of $600\,$mV, i.e. the one of Kittel's experiment. This suggests that the discrepancy observed between the power measured in this experiment and the predictions of conventional FED could be indeed attributed to the contribution of electrons.

Based on this last result, we develop now a simple model to describe the values for the current and heat flux observed in recent experiments~\cite{Kittel,Reddy}. Based on the results shown in Fig.~\ref{Fig_2}, we can neglect the role of phonons. As far as the photonic flux is concerned, we exploit the so-called proximity approximation (PA), already employed in the theoretical simulations described in Refs.~\cite{Kittel,Reddy}. According to this approximation, the extreme end of the tip (probe) is cut in planes of finite area parallel to the sample and the heat flux exchanged between the tip and the sample is simply obtained by summing the heat flux between the sample and all these elementary surfaces. For a spherical shape of radius $R$, the net power exchanged between the tip and the sample reads
\begin{equation}
P_{h}^{\text{(ph)}}=2\pi\int_{0}^{R}dr\, r\, J_{h}^{\text{(ph)}}(d+R-\sqrt{R^2-r^2}).
\label{flux_PFA}
\end{equation}
Concerning the electronic contribution, the PA is generally not valid anymore. Based on the theory of scanning tunneling microscopy~\cite{ChenSTM,Binnig1,Binnig2,Binnig3} we proceed differently. We model the electronic flux and current between tip and sample by taking the plane--plane result and multiplying it by $\pi r_\text{Au}^2$, where $r_\text{Au}=1.35\,$\AA\,represents the radius of a gold atom. This simple approach is validated by the verification that when approaching contact it gives results in agreement with the quantum of electrical conductance $G_0=2e^2/\hbar$. Another very relevant issue is the choice of the barrier height $V(d)$. This is typically determined in any SThM experiment by looking at the rate of exponential decay of the observed current with respect to the distance~\cite{ChenSTM}. In both experiments mentioned above~\cite{Kittel,Reddy}, the authors obtain values for this barrier between 1\,eV and 2.5\,eV, far below the theoretical value (4.7\,eV) expected for ideally clean and bulk samples. As a consequence, since the theoretical results depend (as expected) strongly on the value of the potential barrier~\cite{SupplMat}, for these simulations we replace our model $V(d)$ with values directly taken from the experimental data under scrutiny. Unfortunately, the results presented in Ref.~\cite{Kittel} are associated with different barrier heights, making the theory-experiment comparison more challenging. As a consequence, we compare our theoretical results with more recent experimental results~\cite{NewExp}, obtained in a different temperature configuration ($T_\text{probe}=295\,$K and $T_\text{sample}=195\,$K) and corresponding to a given barrier height. We have fitted the data for the current [show in the inset of Fig.~\ref{Fig_exp}(a)] with either a $\exp[-2\kappa d]$ or a $\exp[-2\kappa (d-d_0)]$ dependence in order to take into account an error on the determination of the distance. We find a value of the barrier height close to 1\,eV and an error bar on the distance of the order of 1\,\AA. In the inset of Fig.~\ref{Fig_exp}(a), we present our theoretical predictions for the current (the shaded region is the one between our theoretical results for $d$ and $d+1\,$\AA), along with experimental data, showing a good agreement. We remind that in this experiment a bias voltage of 600\,mV is applied. The main part of Fig.~\ref{Fig_exp}(a) shows instead the conductance for the same configuration. Albeit the absence of a good agreement between theoretical and experimental results, we confirm indeed the existence of a strong deviation (amplification) with respect to FED results, due to the electronic contribution. We now focus on the experimental results of Reddy and collaborators, and refer in particular to Fig.~2(c) of Ref.~\cite{Reddy}, corresponding to the cleanest sample and smallest temperature difference ($T_\text{probe}=343\,$K and $T_\text{sample}=303\,$K). For the barrier height we take the value of 1.7\,eV given by the authors and we keep our theoretical error bar of 1\,\AA. Concerning the current, we are able to reproduce the values measured by the authors by considering an applied bias voltage of 8\,mV, much smaller than the one of the experiment discussed above. Concerning the current, our model gives a good agreement with experimental results, as shown in the inset of Fig.~\ref{Fig_exp}(b). Concerning the flux, we observe a strong reduction with respect to the other experiment. This is due to a much lower value of the electronic flux, which is in turn due to a two-order-of-magnitude lower applied voltage. Moreover, at the smallest value of the distance shown in Ref.~\cite{Reddy} (approximately 2\,\AA), our theoretical estimate of the conductance is between 0.5\,nW$\cdot$K$^{-1}$ and 2.6\,nW$\cdot$K$^{-1}$, in good agreement with the upper boundary of 2.5\,nW$\cdot$K$^{-1}$ claimed by the authors.

In conclusion, we have introduced the first theoretical framework to investigate individually the contribution of all channels to the heat transfer between two solids near the physical contact and analyzed the role played by the different energy carriers on the heat transfer between two metals at subnanometric distance. We have predicted a giant heat transfer before contact and demonstrated its electronic origin. We have highlighted the strong tunability of heat flux in extreme near-field regime by applying an external bias voltage. These effects could be used to develop novel strategies of thermal management at this scale. They also could be exploited in the field of heat-assisted magnetic recording as well as in energy-conversion technology.

Further theoretical developments should lead to a refinement of comparison of predictions with the experimental results. Among them are a more precise modeling of electron tunneling, a self-consistent calculation of the potential barrier for a given geometric configuration, as well as a more precise description of optical, electronic and mechanical properties.

\begin{acknowledgments}
R.~M. and P.~B.-A. acknowledge discussions with A.~W. Rodriguez. The authors acknowledge financial support by the DAAD and Partenariat Hubert Curien Procope Program (project 57388963).
\end{acknowledgments}

\end{document}